\documentclass[twocolumn]{aastex631}

\def\spose#1{\hbox to 0pt{#1\hss}}
\def\simlt{\mathrel{\spose{\lower 3pt\hbox{$\mathchar"218$}}
     \raise 2.0pt\hbox{$\mathchar"13C$}}}
\def\simgt{\mathrel{\spose{\lower 3pt\hbox{$\mathchar"218$}}
     \raise 2.0pt\hbox{$\mathchar"13E$}}}

\shorttitle{Interrogative Paper Titles}
\shortauthors{Stern \& Grefenstette}

\begin{document}

\title{Do Papers with Titles Ending in a Question Mark Usually Have the Answer ``No''?}


\author{Daniel Stern}

\author{Brian Grefenstette}
\affiliation{The Institute for Interrogative Astrophysics, Pasadena, CA 91109, USA}

\begin{abstract}
Yes.
\end{abstract}

\keywords{April Fools Day Papers (4/1)}

\section{Introduction}
\label{sec:intro}

With almost clock-like precision, astronomers post approximately one paper per day to the arXiv with a title ending in a question mark.  This begs several questions:  Why do they do this?  What's the purpose?  What hidden parameters drive the variations in the frequency? And, of course, these papers confront the long-standing hypothesis: papers with titles ending in a question mark have the answer "no".\footnote{This hypothesis has been aggrandized into a law in other manifestations, such as Betteridge's Law of Headlines which states: ``Any headline that ends in a question mark can be answered by the word no.''  The maxim has also had other incarnations including ``Davis's Law'', apparently a subset of Murphy's Law though the history remains veiled.} 

This paper seeks to provide a definitive, albeit lazy and non-exhaustive, answer to that question.

\section{Methodology}
\label{sec:methods}

Using an old school LLM---i.e., the lead author directing carefully crafted queries to the code-creating second author---we extracted titles for the 4207 papers published on the arXiv between 2026 January 1 and 2026 March 18.  From this sample, 55 papers, or 1.3\%\ of the total sample, have titles ending in a question mark.

As a first step, we disqualified those papers where the interrogative title did not seem best tailored to a simple ``yes'' or ``no'' answer.  Examples of deselected paper titles include ``New Horizons in Effective Field Theory?'' \citep{hollands2026} with its quizzical end punctuation and ``'The X-Ray Dot: Exotic Dust or a Late-Stage Little Red Dot?'' \citep{hviding2026}, which enables us to satisfy the recent requirement imposed by the arXiv that any paper addressing extragalactic astrophysics must mention LRDs.  This disqualification step could be challenging at times, as some paper titles elicited audible grunts and and titular responses of ``no....''---e.g., ``O Nature, Where Art Thou?'' \citep{dowker2026} and ``Why do we do astrophysics?'' \citep{hogg2026}. In total, 20 papers were disqualified, leaving a total sample 35 papers with expected binary yes/no titular responses.

We note that several papers had sufficiently convoluted titles and abstracts that these authors questioned their ability to reasonably digest the science below those opening manuscript lines
and assess the answer to the question mark adorning the paper title.
An example of such a paper is ``Does relativistic motion really freeze initially maximal entanglement?'' \citep{li2026} which investigates the relativistic dynamics of quantum entanglement in a four-qubit cluster (CL$_4$) state using a fully operational Unruh-DeWitt detector framework.

\section{Results}
\label{sec:results}

From the final sample of 35 papers with binary yes/no titular responses, we found a plurality of
14 papers with likely ``no'' responses,  compared to 10 papers with likely ``maybe'' responses, 9 papers with likely ``yes'' responses, one ``probably'' paper, and one ``uncertain'' paper.

\section{Conclusions}
\label{sec:conclusions}

Based on this limited sample, we quantitatively support the hypothesis that paper titles ending with a question mark and expecting a yes/no response do indeed strongly favor the negative response.

\eject
\bibliography{bibliography}{}
\bibliographystyle{aasjournal}

\end{document}